# Normalisation de terminologies multilingues pour les TICE : techniques et enjeux

**Henri Hudrisier**
Laboratoire Paragraphe Université Paris 8, Liaison A de l'AUF à l'ISO SC36
Henri.hudrisier@wanadoo.fr

**Mokhtar Ben Henda**
ISD, Université La Manouba & ISIC, Université Bordeaux 3, *Convener* WG1 à l'ISO SC36
Mokhtar.benhenda@u-bordeaux3.fr

**Résumé :**

La normalisation de la terminologie et de la lexicographie est un enjeu fondamental qui devient de plus en plus crucial à l'ère de la mondialisation multilingue et pour ce qui nous concerne à l'époque de la terminotique et de la traductique. Ces enjeux de la globalisation multilingue mais aussi ceux de l'e-sémantique impactent directement sur les méthodes en normalisation :
- ➢ développement et devenir des normes en « Terminologie et autres ressources langagières et ressources de contenu » (c'est le titre de l'ISO-TC37)
- ➢ développement et devenir de tous les champs normatifs qui développent des terminologies (ou des vocabulaires) le plus souvent multilingues servant de base à leur développement puis de référent à leur usage.

Dans la première partie de notre présentation nous ferons d'abord le point sur les aspects normatifs de la normalisation en terminologie et surtout en terminotique.
Dans une deuxième partie nous ferons un panorama rapide des chantiers terminologiques en normalisation et de leurs raisons d'être
Avec la troisième partie nous développerons le point particulier des TICE. Nous insisterons sur notre implication dans ce chantier, nos hypothèses et convictions de méthodes. Nous exposerons nos développements théoriques et techniques en cours et nous conclurons sur nos besoins de collaboration avec votre communauté académique.

**Mots clefs :**

Terminologie multilingue, normalisation, e-Learning, TICE, SC36

**Abstract :**

Terminology and lexicography standardization is a fundamental issue that is becoming increasingly important in the era of the multilingual globalization and particularly of, from our standpoint, the era of terminotics and translation. The challenges of the multilingual globalization and e-semantics directly impact on standardization methods:
- ➢ Development and perspectives of standards for "Terminology and other language and content resources" (the title of ISO-TC37)
- ➢ Development and future of all standardization fields that develop terminology (or vocabulary) most often multilingual, serving as the basis for their development and acting as a reference to their use.

In the first part of our presentation we will first point on the normative aspects of standardization in terminology and especially terminotics.

In a second part, we will present a brief overview of terminology standardization projects and their rationale,

Within the third part, we will develop the specific issue of ICTE. We will focus on our involvement in this field, on our assumptions and values of methods. We will set out our theoretical and technical developments underway and will conclude with our needs for collaboration with your academic community.

المستخلص:

تقنين المصطلحات والمعاجم قضية أساسية تزداد أهمية في عصر العولمة والتعددية اللغوية وتحديدا في عصر المصطلحية المحوسبة (terminotics) والترجمة الآلية. فتحديات العولمة والتعددية اللغوية من ناحية وكذلك مقتضيات الدلالية الإلكترونية (e-semantics) تؤثر تأثيرا مباشرا على أساليب التقنين من حيث :

- تطوير مواصفات "المصطلحية وغيرها من الموارد اللغوية وموارد المحتوى" (وهو عنوان اللجنة التقنية TC37) والمراهنة عليها،
- تطوير كافة المجالات التي تنمي تقنين المصطلحات (أو المفردات) غالبا بلغات متعددة، والتي تُستخدَم كأساس لتنميتها، وتكون بمثابة المرجعية لاستخدامها والمراهنة عليها.

سوف نتعرض في الجزء الأول من ورقتنا إلى الجوانب المعيارية لتقنين المصطلحات وخصوصا المحوسبة منها (terminotics).

كما سنقدم في الجزء الثاني لمحة موجزة عن المشاريع المتعلقة بتقنين المصطلحات ومبرراتها،

ثم نختم في الجزء الثالث بالتعرض إلى المسألة الخاصة بتقنات المعلومات والاتصال في مجال التعليم لنركز من خلالها على مشاركتنا في هذا الميدان المتعلق بالمصطلحية و على افتراضاتنا واختياراتنا المنهجية في تطوير نظرياتنا حول مبادئها وأساليب عملنا الميدانية من أجل تطويرها. وسنختم مداخلتنا بطرح توجهاتنا واحتياجاتنا لبناء شراكة عملية مع المجموعات الأكاديمية التي ترغب في ذلك.

الكلمات المفاتيح:

المصطلحية ـ التعددية اللغوية ـ المعايير والمواصفات ـ التعليم عن بعد ـ اللجنة الفرعية 36 ـ

# 1. Introduction

Nous sommes heureux de pouvoir participer à cette manifestation scientifique sur un axe que nous avons toujours considéré d'actualité et porteur d'innovation. Depuis quelques années, nous avons été amenés à nous impliquer dans des chantiers internationaux de normalisation de la terminologie, sans pour autant prétendre être des terminologues, ni des traducteurs ou de linguistes de profession. Nous arrivons à ce domaine stratégique par le biais de deux pistes convergentes : d'une part notre spécialité en sciences de l'information et de la communication qui fait largement appel à la terminologie dans ses mécanismes de référencement des ressources, et d'autre part notre métier d'enseignants-chercheurs concernés directement par les questions de la pédagogie universitaire et impliqués dans les nouveaux modes d'enseignement à distance par les TICE. La terminologie multilingue, cumulative, interopérable et globalisable parce que normalisée, devient un enjeu fondamental qui attire de plus en plus d'acteurs toutes disciplines confondues. A l'ère de la mondialisation et du tout numérique, la capacité de traductique, d'interactivité interlinguistique et inter-écriture, la terminotique aussi bien sûr constituent les préalables indispensables de la construction d'un véritable Web-sémantique.

Dans notre communication nous développerons trois champs de recherche et de pratique qui sont pour nous étroitement liées :

- Celui de la normalisation : en fait la condition indispensable pour que continuent à se développer des technologies multimédia interopérables et intercompatibles à l'échelle planétaire d'Internet,

- Celui des TICE (Technologie de l'information et de la communication pour l'enseignement) et des TIC et de leur normalisation,

- Celui de la terminotique normalisée qui constitue évidement le sujet principal de cette intervention.

Nous examinerons les principes de la construction de vocabulaires pour l'*e-Learning* conformément à des normes et des standards internationaux en vigueur comme le modèle TMF (*Terminological Markup Framework*). Nous aborderons le débat des méthodes pour la construction de terminologies normalisées entre l'approche onomasiologique et sémasiologique et nous mettrons l'accent sur les aspects normatifs nécessaires à l'appui de l'interopérabilité sémantique des dispositifs éducatifs. Le cadre de l'ISO/CEI JTC1 SC36 constituera notre espace de référence, très marqué par les enjeux et les débats très accentués de la normalisation dans le domaine du vocabulaire et de la terminologie.

## 2. Les aspects de la normalisation en terminologie : un cadrage conceptuel

La question du vocabulaire est stratégique à plus d'un titre. La terminologie normalisée est en effet l'un des fondements de ce qui est en cours de développement dans les réseaux sémantiques qui caractérisent le Web 3.0. Dans tous les domaines du savoir humain, développer une terminologie normalisée devient indispensable et constitue un instrument essentiel du travail des techniciens du référencement des ressources d'information. Le vocabulaire est aussi important pour les développeurs de contenus. Ceux-ci sont de plus en plus associés au référencement de leurs productions scientifiques par des valeurs terminologiques qui doivent répondre à des valeurs conceptuelles et sémantiques communes. Pour faire face à l'ouverture exponentielle des réseaux et des systèmes d'information, et aux risques de la dispersion des ressources, la normalisation du champ terminologique agit encore une fois en tant qu'agent fédérateur d'initiatives et de modèles.

Nous commencerons notre exposé par définir les grandes lignes du contour épistémologique de la terminologie avant d'évoluer vers une approche plutôt analytique de ce domaine, de ses rapports avec l'activité de normalisation et des enjeux qu'ils peuvent avoir sur la diversité des langues et des cultures.

### 2.1. La terminologie entre le vocabulaire et le champ notionnel des concepts

Il n'est pas inutile, avant d'entrer dans les détails du champ terminologique, d'évoquer en quoi consiste la terminologie, quels sont ses caractéristiques et les courants de pensées qui l'orientent. À moins d'être expert en linguistique, on a souvent tendance à confondre la terminologie avec le vocabulaire d'une part et la terminologie avec la lexicographie ou l'ontologie d'autre part.

### 2.1.1. Définitions

Si l'on part de la définition que la terminologie est l'étude, à la fois, des vocabulaires de spécialité, et de vocabulaires qui peuvent aussi relever de la langue courante ([1]), une première forme de rapport entre « Terminologie » et « Vocabulaire » se dégage. La terminologie est une science et le vocabulaire est l'un de ses champs d'application.

En terminologie, un terme est la combinaison indissociable d'une dénomination (expression linguistique représentant un mot métier) et d'un concept (parfois appelé notion) qui en représente la signification. Un terme est donc un mot dans une langue qui réfère à un concept sous-jacent. Les termes et leurs définitions sont collectés dans des référentiels terminologiques qui conservent les relations qui relient un terme avec les autres. Les relations les plus classiques sont la synonymie, l'antonymie (deux termes de sens opposés), l'hétéronymie (deux termes identiques ayant des sens différents) et la relation « voir aussi » (renvoi d'un terme à un autre). Dans des systèmes d'information hétérogènes, des conflits de données peuvent exister entre des domaines différents (administration, santé, commerce etc.). Les cas les plus fréquents de ce genre de conflits sont la présence de synonymes (deux mots distincts ayant le même sens), d'homonymes (des mots de même graphie et de même prononciation mais de sens différents), d'hyperonymes et d'hyponymes (désignant des niveaux de hiérarchie par le concept « plus général » ou « plus spécifique »). Dans un système d'information, les référentiels terminologiques prennent souvent plusieurs formes : des nomenclatures, des thésaurus, des dictionnaires de données, des référentiels métiers etc. Ils facilitent la mise en relation d'un terme par rapport à l'autre et permettent ainsi de circuler d'un document à l'autre par le suivi des liens qu'ils entretiennent.

---

[1] Le terme « Voiture » relève à la fois de la terminologie des transports et de la langue courante. Mais on voit bien par exemple qu'une terminologie des transports ferroviaires lui donnera une acception très particulière : celui d'un wagon spécifiquement dédié au transport des voyageurs.

Le « Concept » ou la « Notion » est à un niveau d'abstraction plus élevé ([2]) que le « Terme » ou le « Mot ». Il s'agit, pour le linguiste, du rapport entre un « signifié » et un « signifiant » ou un « signifié » et un « référent » (Pottier, 1992). Selon la définition de l'Académie française, le concept est une « *vue de l'esprit, une idée qu'on se fait d'une chose en la détachant de son objet réel* ». Il s'agit, de façon générale, d'une idée abstraite ou d'un concept mental qui se distingue aussi bien de la chose représentée que du mot ou de l'énoncé verbal qui la représente. Teresa Cabré, directrice de *l'Institut Universitari de Lingüística Aplicada* de l'Université *Pompeu Fabra* de Barcelona, le définit dans ces termes : « *Les concepts, ou représentations mentales des objets, sont le fruit du choix des caractères pertinents qui définissent une classe d'objets et non pas des objets individuels* » (Cabré, 1998 & 1999). Selon la norme ISO 704:2000 ([3]), en langage naturel, les concepts peuvent prendre la forme de termes, d'appellations, de définitions ou de toutes autres formes linguistiques. En langage artificiel, ils peuvent prendre la forme de codes ou des formules dans les graphiques. Ils peuvent prendre la forme d'icônes, d'images, de textes, de schémas ou d'autres représentations graphiques. Les concepts peuvent aussi être exprimés avec le corps humain comme ils le sont aussi dans la langue des signes, des expressions faciales ou des mouvements du corps. Le « concept » est une notion fondamentale partagée par la terminologie et l'ontologie.

L'ontologie n'est pas facile à définir de façon générale et définitive car elle prend racine dans des contextes très différents comme la philosophie, la psychologie, la linguistique, l'informatique ou l'intelligence artificielle. On pourrait toutefois partir de la définition très courante de Gruber qui identifie l'ontologie comme une spécification explicite et formelle d'une conceptualisation partagée (Gruber, 1993). La conceptualisation prend forme dans le vocabulaire formalisé de concepts, de leurs relations ainsi que des hypothèses propres à un domaine particulier. L'ontologie fournit une base de conceptualisation partagée sur la base d'une compréhension commune d'un ou de plusieurs domaines par une ou plusieurs communautés. À la différence d'un vocabulaire, donc des mots ou vocables, l'ontologie cherche à représenter le sens des concepts et les relations qui les lient dans un réseau sémantique. En effet, les concepts n'existent pas comme unités isolées de la pensée. Ils sont toujours reliés les uns aux autres dans ce qui est communément appelé un « système conceptuel ». Notre système de pensée crée des relations entre les concepts et les affine constamment pour déterminer si ces relations sont officiellement reconnues ou non. Dans l'organisation d'un schéma de concepts dans une discipline quelconque, il est donc nécessaire de garder à l'esprit le domaine de la connaissance qui a donné naissance au concept et d'examiner les attentes et les objectifs des utilisateurs concernés par cette discipline. Si l'on part de la définition de Rastier (Rastier, 1995), « *toute discipline scientifique a une fonction ontologique* » à partir du moment où elle produit son propre « système conceptuel » à travers une terminologie spécialisée.

---

[2] On notera que sur un plan philosophique, la définition de « Concept » a été conditionnée par plusieurs courants de pensées et a pris plusieurs dénominations comme « Notion complète » chez Leibnitz, « entité mentale » chez les philosophes médiévaux, « entité abstraite » chez les philosophes du XXème Siècle comme Carnap. La définition de « Concept » comme signification d'un terme, au sens d'intension ou de dénotation a été même remis en cause par Hilary Putnam dans son ouvrage « The meaning of "meaning" », in Mind, Language and Reality, Cambridge University Press, 1975, p.218 à 227, traduit par Pascal Ludwig dans « Le langage », Flammarion (GF Corpus), 1997

[3] Produite par le TC 37 *Terminologie, autres ressources langagières et contenus*, la norme 704 établit les principes fondamentaux et les méthodes pour préparer et compiler des terminologies. Elle s'applique aux travaux terminologiques effectués dans des domaines scientifiques, technologiques, industriels, administratifs, ainsi que dans les autres domaines de la connaissance.

*Normalisation de terminologies multilingues pour les TICE : techniques et enjeux. Henri Hudrisier & Mokhtar Ben Henda. Colloque terminologique de Sousse 2009*  5

Depuis l'avènement du Web sémantique, les ontologies ont connu un regain d'intérêt traduit par une prolifération d'ontologies électroniques disponibles dans quantité de domaines. L'hétérogénéité de ces ontologies spécialisées a posé un problème d'interopérabilité entre systèmes d'information s'appuyant sur des ontologies variées[4]. Les techniques d'intégration de l'information sont aujourd'hui l'une des solutions qui proposent des vues unifiées sur des sources ontologiques locales grâce à des schémas d'ontologies globales (Benhlima & Chiadmi, 2006).

La sémantique des réseaux est largement définie dans les relations entre les termes constituant une ontologie. Il faudrait rappeler ici que la sémantique se distingue de la terminologie par la fait qu'elle se préoccupe de la relation entre la dénomination et le signifié, alors que la terminologie s'intéresse d'abord à la relation entre l'objet réel et la notion qui le représente. Les relations entre deux termes représentant des concepts sont généralement associées à l'interprétation humaine qui en est faite. Mais, dans les réseaux sémantiques électroniques actuels, une sémantique formelle peut permettre des calculs automatiques pour vérifier si une cohérence existe entre des informations inscrites pour décrire une connaissance. Des relations comme "plus général que" ou "plus spécifique que" peuvent être formellement calculée en comparant les propriétés de différents termes. Cette relation du général au spécifique est la plus utilisée pour décrire des liens terminologiques. Elle suppose l'existence d'un terme « sommet » pour chaque famille (Fig.1). D'un point de vue représentation, les connaissances et leurs relations sont désormais classées par taxonomies, classifications ou thesaurus. Ils prennent forme de graphes conceptuels ou de réseaux sémantiques (Fig.1).

---

[4] Le W3C a notamment proposé un cadre d'interopérabilité standardisé : OWL, Ontological Web Language, acronyme dont ils soulignent le sens anglais : la chouette emblème d'Athéna, déesse de la pensée, des arts et des sciences.

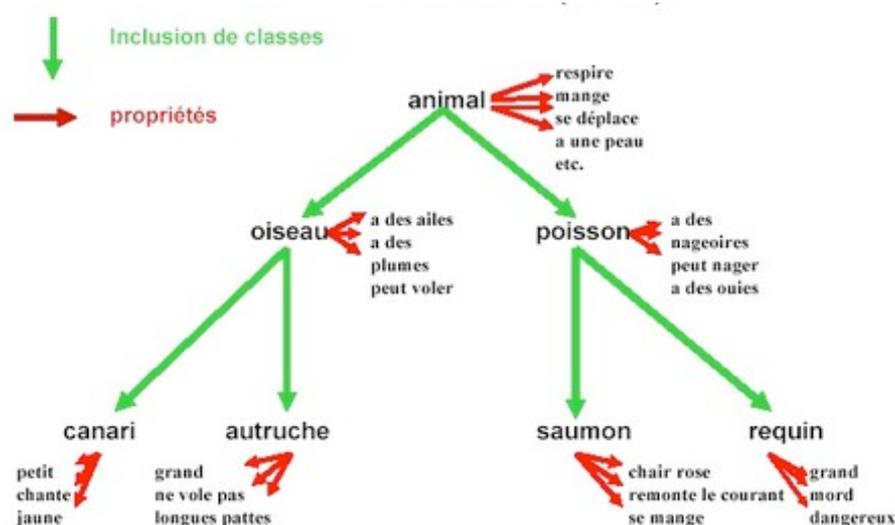

Fig. 1 : Exemple d'un graphe conceptuel (réseau sémantique). (Source : Collin & Quillian, 1969)

### 2.1.2. La terminologie : une science transversale

Comme définition générale du domaine terminologique, le Grand Larousse Encyclopédique définit la terminologie comme un « *ensemble de termes rigoureusement définis d'une technique, d'un domaine particulier de l'activité humaine* ». Le Petit Robert 1, en donne deux acceptions différentes : « *Ensemble des mots techniques appartenant à une science, un art, à un chercheur ou à un groupe de chercheurs* » ; puis « *Étude systématique des termes ou mots ou syntagmes spéciaux servant à dénommer classes d'objets et concepts[5]* ». La Conférence des Services de traduction des États européens précise que la « *terminologie désigne en premier lieu le vocabulaire des langues de spécialité (vocabulaire spécialisé) ; il désigne également la science qui étudie, d'une part, les notions et leurs dénominations dans le cadre des vocabulaires spécialisés (étude théorique) et, d'autre part, les méthodes propres au travail terminologique* » (CST, 2002).

La terminologie est aussi largement abordée par la littérature scientifique. Même si un examen des définitions proposées par les auteurs de référence dans ce domaine ne suffirait pas pour comprendre toutes les subtilités de ce domaine complexe, il serait intéressant d'en donner un court aperçu en guise d'une revue sommaire de la littérature de ce domaine scientifique. Pour Alain Rey, linguiste et lexicographe français, auteur d'un « Que Sais-je ? » sur la terminologie, « *La terminologie est l'étude systématique des termes servant à dénommer classes d'objets et concepts ; les principes généraux qui président à cette étude* » (Rey, 1979). Pour Alain Rey, Les principaux critères que la terminologie met en œuvre sont donc ceux d'une classification structurée. Une pareille conception est reproduite par la linguiste russe Olga Sergeevna Ahmanova (Ahmanova, 1966, cité dans Slodzian, 1993 et rapportée par Gaudin, 2002) : « *La terminologie d'un domaine scientifique, ce n'est pas une liste de termes, mais l'expression sémiologique d'un système conceptuel* ». Pour Guy Rondeau, titulaire de la chaire de terminologie à l'Université Laval, « *La terminologie est l'art de repérer, d'analyser et, au besoin, de créer le vocabulaire pour une technique donnée, dans une situation concrète de fonctionnement*

---
[5] Soulignons que cette deuxième acception apparaît lors de la refonte du Petit Robert en 1983. Elle était complètement absente dans la première édition de 1967.

*de façon à répondre aux besoins de l'usager* » (Rondeau, 1984). Elle fournirait ainsi les termes propres à une activité, en les structurant pour en faciliter l'utilisation.

Dans le contexte français, Loïc Depecker, linguiste terminologue, président de la Société française de terminologie, introduit le champ terminologique, ses principes et ses enjeux à la fin du XXème siècle dans son ouvrage « *Les enjeux de la terminologie* » (Depecker, 1991). Henri Béjoint et Philippe Thoiron (Béjoint & Thoiron, 2000), proposent dans leur ouvrage « *Le sens en terminologie* » un large panorama des différentes questions de sémantique auxquelles conduit la pratique de la terminologie. Michel Foucault, philosophe français établit dans son ouvrage « *Les mots et les choses* » un cadre idéologique et épistémologique des vocabulaires techniques et scientifiques (Foucault, 1966). Daniel Gouadec, Directeur du Centre de formation de traducteurs-localisateurs, terminologues et rédacteurs Université Rennes 2 fait le point, dans son ouvrage « *Terminologie et terminotique* », sur les différentes problématiques liées à la terminologie et à la traduction (Gouadec, 1993).

Tous ces ouvrages de référence, ont un point commun : pour tous ces auteurs, la terminologie, comme une science complexe, prend sources dans presque toutes les disciplines humaines scientifiques et techniques. Elle se trouve au croisement de toutes les disciplines ayant un rapport avec l'information et la communication, la traduction, la rédaction technique, les technologies de l'information et le traitement automatisé du langage. Elle est l'instrument par lequel ces disciplines s'expriment dans un langage et des procédures tout du moins communes. De quelles procédures la terminologie est tributaire ? Quel débat nourrit les tendances procédurales de définition, création et entretien des bases terminologiques ?

### 2.1.3. Le débat de méthodes : entre approche onomasiologique et approche sémasiologique

La collecte des termes (qui est une activité intellectuelle très ancienne) s'organise naturellement selon deux démarches de directions inverses mais synergiques : *sémasiologique* et *onomasiologique*.

1. Sémasiologique : Pour des rédacteurs monolingues, des traducteurs, dans une certaine mesure pour les lexicographes (notamment le rédacteur de dictionnaires de langue ou des dictionnaires bilingues, il est fondamental de partir des termes, des mots ou des expressions (syntagmes) pour pointer sur leur signification (des concepts et leur définition).

2. Onomasiologique : pour des documentalistes, pour des spécialistes du *knowledge management*, des ingénieurs ou des chercheurs analysant des processus techniques ou un phénomène scientifique il est nécessaire d'avoir une démarche de direction opposée. Il faut partir des concepts, que l'on organise en système de concepts (avec des graphes de relations : génériques, partitives, relationnelles), pour pointer sur un 2$^e$ niveau sur des jargons ou des langues de spécialité scientifiques ou techniques (et ce, dans autant de langues –multilinguisme- que nécessaire).

Pour ce qui est de l'aspect humain de l'activité terminologique ou lexicographique, la synergie entre les 2 directions de méthode est la règle. Toutes les terminologies se réalisent en liant les 2 méthodes en synergie dialectique.

Aujourd'hui la communauté des terminologues mais aussi celle des lexicographes se sont entendus pour normaliser une démarche unique. Celle-ci est onomasiologique et ce pour plusieurs raisons.

Une banque terminologique sémasiologique connaît une explosion des liens de relations, surtout si elle est multilingue. En effet, si on construit des banques multilingues en partant des termes on rencontre les plus grandes difficultés à apparier *in fine* les structures conceptuelles, car leur nombre croît exponentiellement par rapport au nombre des termes. De ce fait cette méthode est quasi-incompatible avec la logique informatique.

Notons aussi que l'univers des relations conceptuelles ainsi créé perd toute utilité du fait de son immensité et de son extrême complexité.

*A contrario*, la démarche onomasiologique, découle hiérarchiquement des concepts ce qui s'accorde parfaitement avec la logique du calcul informatique. Sa relative simplicité d'organisation hiérarchique des concepts (génériques et partitifs), lui permet, au contraire de la démarche sémasiologique d'être parfaitement utilisable pour modéliser des référentiels multilingues et multiculturels.

## 2.2. Terminologie et normalisation

La terminologie est en fait une discipline ancienne. Élisabeth Blanchon, professeur à l'Université de Paris Diderot, confirme qu'on peut même attester de ses origines grecques (Blanchon, 1977). Toutefois, la rencontre du domaine de la terminologie avec celui des normes est plus récente.

Notons cependant que dès l'apparition de l'écriture (en Mésopotamie ou en Égypte) se met en place ce que les spécialistes dénomment la « littérature de listes ». En effet, bien avant qu'elle serve à raconter des histoires, relater des faits ou définir des lois, les premiers usages attestés de l'écriture sont des nomenclatures, des inventaires, des listes d'objets nécessaires pour un rituel ou une préparation artisanale ou domestique, mais aussi des généalogies, des listes de rois.... Avec ces premiers écrits, on touche déjà aux fondements d'une activité terminologique, et parallèlement on sait que ces listes, ces nomenclatures ont par définition un effet normatif. L'édition de textes scientifiques, techniques ou historique, tant à Rome qu'ensuite avec l'imprimerie aura aussi cet aspect normatif. Les almanachs, les ouvrages techniques ou scientifiques qu'accompagne une activité classificatrice permanente des savants (Linné par exemple) contribue très tôt (surtout en zoologie et en botanique) à fixer des terminologies rationnellement organisées mettant en œuvre une syntaxe unique et signifiante de la dénomination latine savante des types d'être vivants ([6]). De très nombreux savants ne dédaignant pas

---

[6] En nomenclature biologique, on nomme rangs taxinomiques les niveaux hiérarchiques de la classification scientifique du monde vivant, qui du règne à l'espèce, forment les étages de la pyramide accueillant les taxons de la systématique d'un groupe donné d'animaux, de plantes, de champignons, de protistes, de bactéries ou

parallèlement de mettre en relation dans des dictionnaires cette unique terminologie latine du vivant avec les noms d'animaux ou de plantes attestestés dans les langues vulgaires (et très souvent des variantes dans des parlers locaux). Notons aussi dès le XVIIIe siècle l'importance de l'édition d'encyclopédies (celle d'Alembert et Diderot pour la langue française) qui contribuent à définir, puis à fixer quantité de notions et de dénominations techniques ou scientifiques.

### *2.2.1. Les origines documentaires de la normalisation terminologique*

Pour découvrir plus en détail, les origines de la terminologie, on sera obligé de croiser de nouveau le domaine de la documentation et des bibliothèques en tant que domaine précurseur dans la normalisation des mécanismes et des systèmes d'information.

En effet, la naissance du domaine bibliothéconomique et bibliographique est issue d'une prise de conscience de l'urgence de conserver la masse des connaissances consignées dans les ouvrages spécialisés que les progrès techniques et scientifiques de la société industrielle a pu produire à partir de la fin du XXème siècle. Dans leur ouvrage « *Documentation et philosophie* », (Cotten, Hufschmitt & Varet-Pietri, 2003) affirment que « *dans le même temps où la documentation se constituent comme discipline, définissant ses objectifs, ses moyens et ses méthodes, la terminologie à son tour sort de ses limbes, sous l'impulsion non de linguistes, mais d'ingénieurs confrontés à l'évidence que la normalisation non seulement des produits et des techniques, mais aussi de leurs désignations, est indispensable au développement de l'activité industrielle* ».

Plusieurs arguments corroborent cette hypothèse. D'une part, la diversification des sciences conduit à des interpénétrations de plus en plus complexes entre elles. D'autre part, l'idée d'organiser toutes les connaissances à une échelle mondiale démontre ses limites face à l'augmentation et à la diversification des supports d'information. La classification des connaissances humaines, longtemps soutenue par la documentation et les plans de classifications universelles ([7]), évolue désormais vers une représentation linguistique de l'information. Le système *Uniterm*, fondé en 1953, constitue l'une des premières formes de répertoriage des documents à partir d'un vocabulaire contrôlé (Chaumier, 1992). Il s'agit d'une date importante qui marque les débuts de la construction des thésaurus comme une liste hiérarchisée de termes formant un vocabulaire normalisé et inter-reliés par des relations synonymiques, hiérarchiques et associatives. Les normes et l'informatique sont venues ensuite rapprocher davantage les thésaurus avec les ontologies spécialisées en les faisant converger vers les mêmes problématiques et les mêmes types de solutions pour les résoudre. La convergence vers des systèmes de représentations linguistiques de l'information par les terminologies, les

---

d'archéobactéries. La classification classique propose une hiérarchie codifiée en 7 rangs principaux et 5 rangs secondaires, présentée, dans l'ordre décroissant, de la façon suivante : Monde vivant : règne → embranchement, division ou phylum → classe → ordre → famille → tribu → genre → section → série → espèce → variété → forme. RECOFGE est le sigle mnémotechnique des 7 rangs principaux: Règne/Embranchement/Classe/Ordre/Famille/Genre/Espèce. Selon une syntaxe latine rigoureuse et normalisée, les savants du monde entier se sont entendus depuis plus de deux siècles pour établir une nomenclature normalisée et interopérable pour la dénomination univoque de la totalité du monde vivant.

[7] La CDU : Classification Décimale Universelle, développée en 1895 par Paul Otlet et Henri La Fontaine, puis la CDD : Classification Décimale de Dewey, développée par Melvil Dewey en 1876, sont les deux systèmes les plus connus de la classification du savoir humain.

ontologies, les graphes conceptuels etc. a, cependant, engendré une problématique par rapport aux systèmes de classification numériques par les schémas des classifications numériques comme la CDU ou DEWEY. Les termes et les concepts ne sont pas aussi neutres que les chiffres. L'universalité de l'accès aux chiffres (code des classes) laisse progressivement place aux variations sémantiques des mots, termes et concepts auxquelles il faut associer des valeurs socioculturelles et linguistiques très fortes. Ceci a plus d'incidences dans un cadre de définition de normes internationales pour des usages universels.

### *2.2.2. Tradition savante latine paneuropéenne et panarabisme linguistique*

Permettons nous une diversion que l'ouverture interculturelle et multilingue de ce colloque nous autorise. Une réflexion aussi, que le compagnonnage scientifique et amical des deux auteurs induit de par la permanence de leur questionnement réciproque entre leurs deux cultures (orientale et occidentale) et, au-delà, par la force de leur engagement à préserver la diversité linguistique qui les motive dans la construction de la mondialisation numérique (avec l'ISO notamment).

A ce jour, seule la communauté savante du vivant est restée attachée à préserver la référence d'une nomenclature latine univoque. Le reste du monde savant et technique s'est longtemps senti protégé par le formalisme mathématique de leurs domaines respectifs (notamment pour les sciences de l'ingénieur, la physique ou la chimie). Aujourd'hui, cette univocité supposée des concepts techniques (et même scientifiques) a depuis longtemps explosé du fait des langues dans lesquelles ces concepts se déploient (le plus souvent en concurrence). Les efforts pour « maintenir » ([8]) les sciences et les techniques dans un référentiel commun et partagé sont démesurés. Il est paradoxal de constater l'énergie et les efforts pédagogiques que les terminologistes doivent déployer pour faire entendre qu'un concept dès qu'il se transforme en terme devient dépendant de ses multiplicités d'acception potentielle dans telle ou telle langue naturelle.

Depuis une douzaine de siècle le Monde arabe a la chance de partager en commun une langue sacrée, savante, technique, administrative, médiatique… qui transcende ses disparités de cultures linguistiques dialectales. Une langue de plus, qui par sa notation canonique non accentuée est beaucoup moins dépendante d'une improbable dérive phonétique que la plupart des autres langues (européennes notamment). L'arabe classique, est ainsi une langue prestigieuse associée à la religion et à l'écrit, c'est-à-dire à la culture littéraire, à la science et à la technologie et aux fonctions administratives. Elle est aussi appelée arabe coranique, arabe moderne standard, arabe grammatical ou arabe éloquent. Ces caractéristiques ne l'ont pourtant pas préservé des critiques, souvent fondées, quand à sa modernisation lente sur le plan des terminologies techniques. On attribue même à son statut canonique l'inconvénient d'approfondir les écarts entre les dialectes nationaux et « l'approvisionnement » terminologique dans les langues latines par les emprunts linguistiques et la translitération. On évoque de plus en plus les disparités conceptuelles et terminologiques entre *Maghreb* et *Machreq* arabes inhérentes à cette inertie dans la normalisation et l'unification des termes et concepts parfois dans des domaines littéraires. Un concept comme « encadrement », d'une influence française évidente sur l'arabe littéraire au Maghreb, donne lieu à un terme arabe « Ta'tir » entièrement incompréhensible au Moyen-Orient en raison d'une influence anglo-saxonne dominante dans laquelle le concept d'encadrement n'est point associé à la métaphore de cadre. Les exemples de ce genre sont multiples.

---

[8] Nous employons à dessein ce terme dans son sens francophone et dans celui d'un anglicisme largement répandu.

Notre hypothèse serait dès lors que le Monde arabe, de par sa langue écrite unique, est héritier d'une tradition linguistique particulièrement intéressante pour la mise en cohérence multilingue de nomenclatures normalisées de concepts. Il est de ce fait indispensable de contribuer pour que les avancées techniques et savantes de la terminologie et de la terminotique (l'algorithme est aussi un apport culturel du monde arabe savant) soit accompagnés par le développement de contenus terminologiques arabes conséquents et cohérents construits selon des modèles normalisés qui abordent avec beaucoup d'efficacité les questions des nuances conceptuelles et des terminologies appropriées et interopérables. Le modèle TMF, que nous décrivons plus loin, est très convenable à cette situation particulière grâce à sa méthode onomasiologique qui part d'un accord sur les concepts pour arriver aux termes qui les représentent. Nous y reviendrons avec plus de détails dans les sections à venir.

### 2.2.3. Le TC37, précurseur de la normalisation terminologique

Le vrai développement de la terminologie commence vers le début du XXe Siècle, quand l'IEC (Commission électrotechnique internationale/CEI) entame en 1906, le développement de son vocabulaire VEI (Vocabulaire Électronique International). C'est en 1938 que la première version de ce vocabulaire normalisé a été éditée pour unifier la terminologie électrique mondiale. Cette date coïncide avec une autre date importante, celle de la publication des œuvres les plus marquantes d'Eugen Wüster, ingénieur autrichien, père fondateur de la terminologie moderne. En 1931 Wüster publie sa thèse sur la normalisation internationale de la langue dans les domaines techniques (*Internationale Sprachnormung in der Technik*) dans laquelle il expose les fondements d'une théorie générale de la terminologie. En 1935 il publie son Dictionnaire de la machine outil dont les travaux établissent les bases théoriques de la terminologie moderne. Grâce aux travaux de Wüster, l'ISA (ancêtre de l'ISO), crée en 1936 un Comité technique 37 (TC37) chargé d'élaborer des principes méthodologiques pour harmoniser les terminologies et leur mode de préparation et de présentation.

Le TC37, dirigé à ses débuts par Wüster, marquera l'histoire de la terminologie normalisée dans tous les domaines. La communauté des normalisateurs terminologues et lexicographes adoptent comme seule méthode valide, la démarche onomasiologique (ISO704, Terminologie : principes et méthodes). Ils ont ensuite normalisé un catalogue ouvert de catégorie de données apte à définir des données terminologiques ou lexicographiques (ISO/IEC 12620). Puis ils ont normalisé un cadre commun de mise en œuvre terminotique à même d'assurer l'interopérabilité et la réusabilité des ressources terminologiques indépendamment des diverses banques de données terminologiques. Ce cadre commun, le TMF (*Terminological Markup Framework*) ([9]) que nous décrirons plus loin dans ce document, nécessite bien sûr que ces différentes bases respectent le métamodèle XML TMF, ou exige que les ressources terminologiques soient reformatées selon ce même modèle.

---
[9] Un cadre de mise en œuvre normalisé pour la terminotique, lui-même lié au TML (*Terminological Markup Language*), un langage XML spécialisé (en fait une DTD) pour l'expression des données terminologique. TMF correspond à la norme ISO/IEC 16642

L'intérêt de ces normes du TC37 est qu'elles ouvrent le cadre d'interopérabilité entre différentes bases de données terminologiques. Cette interopérabilité est d'abord assurée par l'identité de méthode et du mode de description. Plusieurs bases de données terminologiques qui s'appuient sur la norme ISO 704 (principes et méthodes) et ISO/IEC 12620 (Catégories de données en terminologie), seront déjà assurées d'avoir constitué des ressources terminologiques qui peuvent (au prix d'efforts informatiques importants mais possibles) être récupérées pour être interopérables entre elles.

Dès lors, il devient clair que ces 10 dernières années les efforts terminotiques ont été très importants et se sont focalisé sur le TMF et l'adaptation des nouvelles versions de ISO/IEC 12620 à ces nouvelles méthodes.

La terminotique n'a réellement bouclé ses méthodes que grâce au progrès de l'information structurée (SGML et surtout XML) qui ont permis d'organiser toutes les banques de données terminotiques (notamment les terminologies multilingues) : aussi bien en permettant de concilier la logique lexicographique et la logique terminologique mais en permettant surtout de concevoir un schéma terminotique unique (TMF) qui permet d'assurer l'interopérabilité de toutes les banques terminologiques ou lexicographique (quelque soit le nombre de langues mises en œuvre dans la même base).

L'interopérabilité des terminologies dans le futur web sémantique exigeait notamment un tel choix normatif. Les mécanismes du TMF, et la normalisation des catégories de données permettent précisément de servir de base à la réalisation modulaire d'ontologies elles même en cours de standardisation par le W3C (langage OWL).

### 2.2.4. Normalisation terminologique : état de l'art

Le site officiel de l'ISO recense aujourd'hui (2009), 178 textes de normes étudiant ou proposant des listes de termes dans tous les secteurs d'activités. On ne cite ci-après que les plus récentes des normes élaborées par le SC2 du TC37 en rapport avec la terminologie :

- ISO 1951:2007 : Présentation/représentation des entrées dans les dictionnaires -- Exigences, recommandations et information
- ISO 10241:1992 : Normes terminologiques internationales -- Élaboration et présentation
- ISO/DIS 10241-1 : Articles terminologiques dans les normes -- Partie 1: Exigences générales et exemples de présentation
- ISO/DIS 10241-2 : Articles terminologiques dans les normes -- Partie 2: Introduction de normes terminologiques internationales dans différents environnements
- ISO 12199:2000 : Mise en ordre alphabétique des données lexicographiques et terminologiques multilingues représentées dans l'alphabet latin ([10])

---

[10] Ce seul problème pour l'alphabet latin correspond à un maquis de pratiques disparates : renvoi ou non en fin de terme entre parenthèse de particules devant des patronymes, renvoi en fin d'alphabet ou en fin de la rubrique d'une lettre de certaines ligatures ou de certains diacritiques (Æ, Ñ…). Prise en compte du radical

- ISO 12615:2004 : Références bibliographiques et indicatifs de source pour les travaux terminologiques
- ISO 12616:2002 : Terminographie axée sur la traduction
- ISO 15188:2001 : Lignes directrices pour la gestion de projets de normalisation terminologique
- ISO 22128:2008 : Produits et services en terminologie -- Aperçu et orientation
- ISO/DIS 23185 : Critères d'évaluation comparative des ressources terminologiques -- Concepts, principes et exigences d'ordre général
- ISO 12620:1995 : Aides informatiques en terminologie- Catégories des éléments de données terminologiques.
- ISO 1087:2000 : Terminologie-Vocabulaire ; nouvelle édition 2000 : Travaux terminologiques-Vocabulaire-Partie 1 : Théorie et application.

Il est tout aussi important de signaler la norme « ISO 704 :2000 : Travail terminologique -- Principes et méthodes » du SC1 du TC37. Cette norme définit les principes fondamentaux et les méthodes de la préparation et de la compilation des terminologies, qu'il s'agisse de créer des terminologies, des vocabulaires, des nomenclatures pour l'usage strict des experts en normalisation ou pour des visées plus larges que ce soit dans un cadre industriel ou linguistique. Elle détermine les principes généraux qui contribuent à la gestion de la formation des désignations et à la formulation des définitions. Elle décrit les liens entre les objets, les concepts et leurs représentations par des terminologies.

La norme ISO 704:2000 définit deux catégories de relations entre les concepts : des relations hiérarchiques et des relations associatives. Les relations hiérarchiques peuvent être du type relation générique ou relation partitive. Le schéma suivant (Fig.2) donne un exemple concret des trois types de relations qui peuvent exister dans un système de concepts. Un concept générique « Instrument d'écriture » se décline en plusieurs concepts spécifiques dans un type de relation hiérarchique du générique au spécifique. Le concept spécifique « crayon » devient à son tour un concept générique et se décline en d'autres concepts spécifiques par une relation du type hiérarchique. Le concept « porte-mine », qui désigne une sous catégorie de crayons, se décompose en concepts partitifs qui constituent les différentes parties du concept générique « porte-mine ». Certains de ces concepts partitifs se décomposent à leur tour en d'autres concepts partitifs d'un niveau inférieur. Par contre, les relations associatives ne sont pas hiérarchiques. Une relation associative peut exister quand il y a un rapport thématique quelconque entre des concepts.

Parmi les relations de dépendance définies par la norme ISO 704:200, on pourrait citer des relations associatives du genre : contenant à contenu ; activité à outil ; cause à effet ; producteur à produit ; durée à outils de mesure ; profession à outils spécifiques de travail etc. Dans le cas de la figure suivante (Fig.2), une relation associative est établie entre la mine du crayon comme une matière d'écriture et le graphite comme un minerai.

d'un terme ou du terme in extenso (en arabe par exemple mais aussi dans d'autres langues) ; mise en ordre des clefs puis du nombre de traits ajoutés à la clef pour des langues idéographiques. Ordre phonétique, ordre de similarité des caractères, ordre de correspondance numérique des caractères, ordre « poétique » des katakana et hiragana japonais correspondant à un poème les contenant tous… Cette diversion est là pour démontrer à quel point il est indispensable que les terminologistes ayant la chance de ne pas appartenir à la culture linguistique (encore) aujourd'hui dominante, se doivent de contribuer à aménager des normes à même de préserver leurs spécificités linguistiques et culturelles.

La plupart des liens relationnels, peuvent établir des « relations » avec un ou d'autres secteur(s) du graphe. En ouvrant des liens vers une autre branche de « l'arbre », le système de concept s'élargit indéfiniment créant le réseau de concepts.

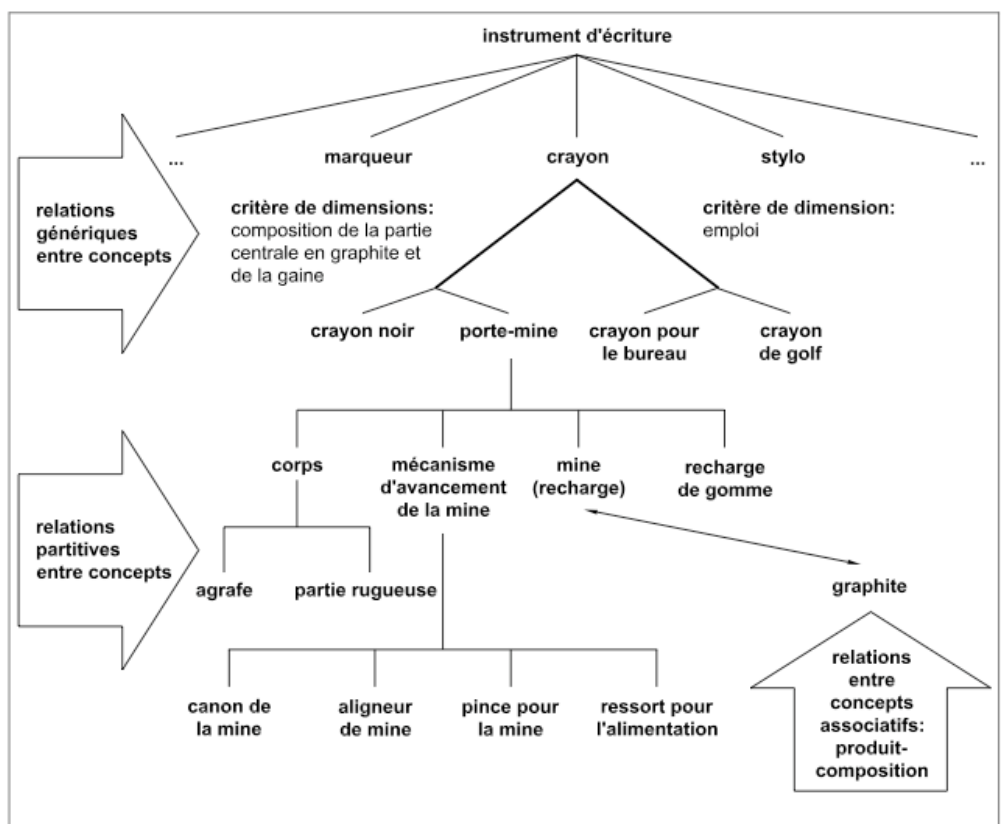

Fig. 2 : Les types de relations dans un système de concepts (Source : la norme ISO 704:2000)

A l'issue de cette large introduction des mécanismes de l'activité terminologique et de ses modèles fonctionnels, nos restons attirés par un aspect singulier, celui de la construction terminologique multilingue normalisée. Avec la mondialisation, le respect de la diversité culturelle et linguistique occupe solidement une position de priorité nationale et internationale. L'internationalisation et la localisation sont désormais les deux faces d'une même pièce. Outils, logiciels et données sont désormais systématiquement soumis à une ouverture systématique à toutes les langues du monde. La terminologie ne se soustrait certainement pas à cette orientation. C'est d'ailleurs l'un de nos chantiers de travail prioritaire comme nous l'avons annoncé dans ce document. Comment sont définies les stratégies d'une terminologie multilingue ? Et quels en sont les enjeux de son développement pour les communautés de pratique ?

## 2.3. Pour un système terminologique « larges langues » et normalisé

Nous proposons ce terme de « larges langues » volontairement accrocheur ([11]), pour bien signifier qu'à l'ère de la mondialisation il n'est plus temps de continuer à croire que des terminologies bi, voire quadrilingue continuent d'être suffisantes pour répondre aux besoins de la communication mondiale. Les technologies, notamment pour les terminologies, sont maintenant susceptibles de s'ouvrir à un nombre extensif de langues (plusieurs dizaines dans un nombre maximum d'écritures). Nous ne prétendons pas que ce genre de projet soit facile à développer. Nous savons même qu'il pose un maximum de problèmes tant conceptuels que techniques que nous tentons d'expliquer ci-après.

---

[11] Construit sur le même patron linguistique que large bande (*broadband*) : largement ouvert sur un grand nombre de fréquences et permettant de hauts débits.

### 2.3.1. L'appréhension de concepts définis en consensus dans une ou plusieurs langues

Dans l'un de nos travaux précédents sur la question des vocabulaires et terminologies e-Learning (Hudrisier & Ben Henda, 2008), nous avons démontré comment l'appréhension de concepts définis en consensus est une nécessité pour des experts travaillant sur la définition d'une norme dans une seule langue. La démarche devient beaucoup plus rigoureuse pour des experts appartenant à des communautés linguistiques disparates pour travailler dans une langue de travail commune ([12]). Mais la conscience collective d'œuvrer pour des perspectives normatives multilingues s'est souvent traduite, sous l'impulsion de ces experts multilingues, en plusieurs textes de normes appuyant la diversité des langues et des cultures.

Dans l'élaboration de normes internationales, il est de ce fait impératif de fonctionner sur la base de concepts définis en consensus pour que la déclinaison de ces concepts en termes dans autant de langues concernées ne soit pas une simple traduction d'une langue source à une langue cible ([13]). C'est uniquement en confrontant les diversités du découpage des concepts, induites par les disparités des cultures et des langues que l'on peut être assuré de pouvoir rédiger ensuite des normes internationales capables de répondre à la multitude des diversités des usages partout dans le monde. C'est surtout par cette approche qu'il sera possible de localiser ces normes par traduction ou adaptation afin que les acteurs nationaux puissent ensuite se les approprier, chacun dans son cadre culturel et linguistique approprié. Cette méthode s'appuie sur des théories bien connues du domaine terminologique, en l'occurrence celles des méthodes onomasiologiques et sémasiologiques dans la construction des vocabulaires normalisés. Nous aborderons ces théories quand nous évoquerons plus tard les mécanismes de fonctionnement du modèle TMF (*Terminological Markup Framework*) pour la construction de terminologies multilingues.

Les terminologies « larges langues » auront aussi plusieurs effets sur le déroulement des travaux des instances de normalisation. Elles permettent d'aider les experts à rédiger ou à traduire de façon cohérente les normes du domaine dans leurs propres langues. Une terminologie normalisée, fondée sur un consensus de concepts, permet de vérifier que l'on parle bien des mêmes choses (des mêmes concepts, et non pas seulement des mêmes mots), des mêmes produits, des mêmes composants, des mêmes logiciels, des mêmes services, des mêmes processus, des mêmes ressources, des mêmes institutions, des mêmes acteurs, des mêmes rôles, etc. Ces terminologies doivent, par contre, être vérifiées et validées avec les réseaux d'experts auxquels chaque expert international est lié, notamment la maîtrise en « miroir » avec les experts référents dans leurs délégations nationales ou Liaisons ([14]).

---

[12] À l'ISO, l'anglais, le français et le russe sont les langues officielles de travail (ça ne veut pas dire langue réellement admises dans les sous instances. Au JTC1, l'anglais est la seule langue de travail.

13 On prend souvent comme exemple la désignation des phases successives de l'éducation qui sont loin d'être universelles : école maternelle, école primaire, enseignement secondaire (découpé ou non en collège puis lycée), l'enseignement universitaire ou professionnel correspondent dans d'autres cultures à des regroupements qui peuvent être très différents. On voit bien de ce fait que la terminologie interculturelle et interlinguistique est bien autre chose qu'une correspondance terme à terme. On entrevoit donc sur cet exemple apparemment trivial la difficulté à construire réellement un référentiel sémantique commun indépendant des langues et des cultures.

Ces terminologies ont également un rôle prospectif en constituant des référentiels qui pourraient aider les futurs utilisateurs des normes à avoir la même compréhension des mêmes concepts. La difficulté du second objectif réside en ce que ces utilisateurs finaux ne maîtrisent pas nécessairement de façon experte la totalité des concepts proposés par les experts internationaux. Ils ne maîtrisent pas non plus les langues de travail des experts internationaux. Cependant, ces utilisateurs finaux sont confrontés à des particularismes culturels, linguistiques, institutionnels qu'il s'agit de préserver dans les situations locales réelles où ils ont besoin de mettre en œuvre les normes considérées.

De là, découle la nécessité et l'urgence de s'entendre sur des méthodes de construction de terminologies multilingues fondées sur des consensus s'établissant sur des concepts. Les référentiels sémantiques et les systèmes de concepts deviennent dès lors indispensables pour permettre l'interconnexion et l'interopérabilité des matériels et des réseaux ainsi que celles des ressources pour qu'elles puissent notamment circuler et interagir malgré la diversité des langues.

Pour systématiser cette appréhension de concepts définis en consensus, les institutions de normalisation conseillent instamment aux animateurs de chaque instance de normalisation ([15]) de créer un groupe de travail spécifique chargé d'élaborer une terminologie du domaine. Notons cependant que tous les experts en normalisation dans tous les métiers qui peuvent être concernés sont diversement sensibles à la nécessité de normaliser les terminologies de leur domaine de spécialité (malgré les normes de l'ISO TC37). De ce fait, les terminologies produites par l'ISO sont méthodologiquement et techniquement très disparates. La coordination interdisciplinaire de terminologies spécialisées n'est pas une question simple. Par contre, la mise en place de terminologies rigoureusement normalisées (pour ce qui est de leur méthode et de leur interopérabilité terminotique) est une voie d'avenir, conforme aux avancées vers un véritable web sémantique. C'est aussi la seule façon connue de mettre en place un environnement TIC permettant de répondre aux défis de la mondialisation.

Une fois ce premier niveau de compréhension des concepts acquis pour les experts, il s'agit qu'ils puissent communiquer à l'écrit comme à l'oral malgré leurs niveaux inégaux de maîtrise de la (ou des) langue(s) de travail. Contrairement aux documents de la communication quotidienne, littéraire ou même diplomatique les documents normatifs (comme les documents juridiques) sont très difficiles à comprendre non pas seulement parce qu'ils sont écrits dans une langue qui peut nous être étrangère, mais parce qu'ils sont écrit dans un style faisant une très large place au références, aux sigles, aux acronymes divers. Comme le langage juridique (et de façon beaucoup plus systématique que les documents scientifiques et techniques) le rédacteur de norme doit utiliser un vocabulaire codifié, rarement explicité même quand il fait référence à des commissions de normalisation autres que celles qui produisent le document. Il faut comprendre ici que les experts en normalisation se réunissent essentiellement à 2 niveaux :

1. Le niveau de leurs institutions nationales ou ils élaborent soit des normes qui restent nationales (donc souvent unilingues sauf le cas des pays multilingues),

2. Le niveau international qui implique que tous les experts délégués à un niveau international travaillent bien sûr au strict niveau de l'ISO (et instituts similaires) mais aussi à leur niveau national (AFNOR, EANOR, INNORPI…) où ils rendent compte, discutent, collaborent avec leur niveau national pour que l'ensemble des communautés nationales d'experts contribue à la construction des normes internationales.

---

[14] Beaucoup de NBs (National Bodies : les Etats) et certaines Liaisons réalisent des terminologies ou des vocabulaires : dans le cas du SC36 on doit citer notamment : la GB, le Canada, la Corée, le Japon, la France, l'Allemagne… Pour ce qui est des Liaisons on peut citer l'AICC, IMS, IEEE qui ont tous des terminologies considérables sur les TIC (et aussi sur les TICE).

[15] Une instance de normalisation est animée par un président (chairman), matériellement assisté par un Secrétariat général (en général soutenu par un NB qui en prend la responsabilité, dans le cas du SC36, c'est le BSI, British Standards Institute). Toutes les sous-instances (les WGs ou RG éventuels), sont elles mêmes animées par des Présidents de WGs ; s'ajoutent à cela des éditeurs de normes, les Présidents des différents NBs (et des Liaisons). L'ensemble constitue une sorte d'état major hiérarchisé qui anime la production d'une norme. Remarquons toutefois que ces animateurs doivent avoir pour soucie constant la recherche du consensus maximum (ce qui les distingue d'un état major militaire ou industriel) et le verdict constant des votes des NBs. Ces animateurs doivent aussi veiller constamment à ne pas être en contradiction avec les instances ISO, IEC, etc… dont ils sont des sous-instances et qui ont sur eux un contrôle lui aussi soumis aux mêmes règles consensuelles et à des processus de vote des États membres.

On comprend de ce fait que ce niveau de travail complique encore les tâches de traductions, de commentaires d'aide à la compréhension de textes en langue étrangère pour ces communautés nationales d'experts.

Pour qui a l'expérience de ce type de commission, il faut avoir le courage d'admettre que des décisions sur des normes quelquefois cruciales pour nos sociétés sont souvent prises par des assemblées d'experts qui pour certains d'entre eux maîtrisent parfaitement la compréhension des textes, mais pour d'autres « ânonnent leur traduction » tout en faisant semblant de comprendre. S'ils « ânonnent » ce n'est pas parce qu'ils ne comprennent pas les mots et les phrases dans lesquels sont rédigés les documents (souvent de l'anglais), mais c'est parce qu'à ce niveau de traduction ou d'interprétation (déjà compliqué pour eux mais qu'ils pourraient surmonter) vient s'ajouter une avalanche de références, de sigles, d'acronymes (dont le pire est quelquefois qu'ils soient traduit dans la propre langue de l'expert lecteur[16]). Là encore, on comprend que les outils d'aide à la traduction et les terminologies électroniques sont indispensables.

Soyons en effet très directs par rapport aux risques supposés de manque à gagner pour les activités professionnelles comme la traduction et l'interprétation. Aussi lourds soient les crédits que nous attribuions à des tâches de traduction et d'interprétation, ils seront toujours très largement insuffisants pour couvrir tous les besoins d'une activité comme la normalisation. De notre point de vue il est clair qu'il faut plaider à nombre de niveau de décision de nos institutions pour que soient fait d'énormes efforts financiers pour impliquer toujours plus de professionnels dans des tâches de traduction ou d'interprétation pour la normalisation, mais :

1. La localisation (car s'en est une) des documents normatifs, doit se réaliser, comme la localisation des logiciels, en faisant coopérer un traducteur professionnel, des outils d'aide à la traduction et un expert normalisateur impliqué dans le processus normatif concerné,

2. Les réunions d'experts sont souvent complexes, diplomatiquement délicates et deviendraient impossible si elles devaient fonctionner dans un environnement de traduction simultanée.

---

[16] L'expert habitué aux sigles cités dans les travaux internationaux (très souvent des sigles anglophones) peut être totalement désorienté quand on lui propose un sigle décliné dans sa propre langue. De façon générale il est très difficile d'appréhender des sigles traduits tant en lecture qu'en écoute de texte étranger, par exemple ITC= TIC ; SMSI=WSIS).

D'autre part le nombre de communautés linguistiques engagées dans ces réunions internationales est tel qu'il serait irréaliste de l'envisager dans un nombre extensif de langue (ou même seulement avec les 3 langues officielles de l'ISO : anglais, français, russe). D'autre part ré-insistons sur le caractère très spécialisé de l'activité qui entraînerait inéluctablement de très nombreux contresens, pire sans doute que la frustration de la non compréhension immédiate compensée (si l'expert est sérieux), par une traduction ultérieure à tête reposée dans laquelle s'impose encore l'importance des aides électroniques.

### *2.3.2. Le modèle TMF comme process technique d'élaboration des normes terminologiques et les conséquences de son appréhension multilingue*

Si nous avons choisi de poursuivre avec le sujet de la diversité linguistique dans la définition des terminologies normalisées, c'est essentiellement pour mettre en avant l'un des points clés de la méthodologie de recherche-action que nous utilisons et que nous voudrions communiquer à des partenaires potentiels dans l'auditoire. Il s'agit de nos projets de définition et de mise en place de terminologies normalisées *'larges langues'* dans le cadre de nos activités dans le programme d'action du SC36. Ces projets sont définis sur des acquis de la normalisation que nous avons cités auparavant. Ils s'inspirent largement de travaux antérieurs accomplis dans ce domaine et de référentiels internationaux confirmés en la matière.

Notons déjà que deux intervenants de ce Colloque de Sousse 2009 (Béchir Boudir et Joseph Tiencheu) ont contribué substantiellement et contribuent encore aujourd'hui aux travaux du SC36 WG1 que nous présentons ici. Ils ont notamment réalisé une modélisation du système de concept de cette terminologie. Béchir Boudir vient de prendre il y a quelque mois la responsabilité de l'animation de l'AFNOR CN36 GE1 (Groupe d'Expert en terminologie) qui est l'instance-miroir du SC36 WG1 (niveau ISO) dont Mokhtar Ben Henda est le « *convener* ».

Les travaux de Laurent ROMARY, directeur de recherche à l'INRIA et chairman du SC2 du TC37, font référence dans ce domaine. Parmi les problèmes qu'il identifie dans la représentation d'une terminologie multilingue on repère une notion importante : celle de « *l'organisation générale de ces données en grandes composantes* » (Romary, 2001). En effet, conformément à notre analyse précédente de deux approches onomasiologique et sémasiologique, deux modèles concourent à l'organisation des données terminologiques multilingues. D'une part, une organisation onomasiologique inspiré du modèle de Wüster qui structure une base terminologique en plusieurs niveaux hiérarchiques allant du niveau de concept vers le niveau de la langue et atteint enfin le niveau du terme (Fig.69).

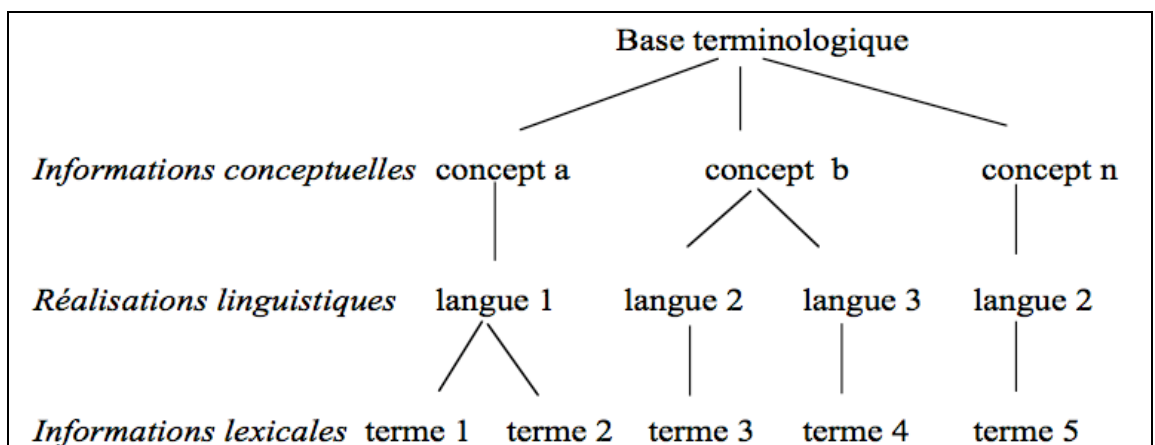
Fig. 3 : L'approche onomasiologique multilingue (Source : Kramer, 2004)

D'autre part, une méthode opposée, utilisée dans les systèmes de traduction automatique s'appuie plutôt sur une description lexicographique. Elle part des termes ou des expressions d'une langue source pour créer des associations avec leurs équivalents dans une langue cible. L'association couvre les termes ou les expressions, leurs définitions ainsi que toutes les informations syntaxiques, morphosyntaxiques et sémantiques qui les caractérisent. La définition des différents types d'informations complémentaires à chaque terme ou expression, comme sa catégorie grammaticale, sa définition, ses conditions d'usage etc. constitue un autre problème pour la mise en correspondance de ces valeurs avec celles issues d'autres bases. Devant cette complexité, l'usage des standards de représentation de ces données était devenu nécessaire. La définition du modèle TMF a été la réponse à cette nécessité.

Le modèle TMF (*Terminological Mark-up Framework*), normalisé ISO 16642:2003, fournit un cadre de représentation des bases de données terminologiques en langage XML. TMF est un schéma structurel innovant qui vient élargir les perspectives du champ terminographique resté longtemps contraint aux modèles d'échanges définis par la norme SGML d'échanges négociés MARTIF (ISO 12200:1999) ([17]). Après le changement de stratégie de l'ISO par rapport à l'inventaire clos des catégories des données (norme ISO 12620:1999) ([18]), une nouvelle famille de normes d'échanges plus génériques est venue renforcer l'enrichissement permanant des types de donnés. Dans leurs étude sur les relations conceptuelles dans le processus de normalisation des échanges de données en terminologie, Laurent Romary, et Marc Van Campenhoudt, chercheur au centre de recherche TERMISTI en Belgique, jugent ce changement comme « *particulièrement*

---

[17] MARTIF : MAchine-Readable Terminology Interchange Format est le fruit de l'évolution des réflexions menées initialement dans le cadre de la TEI (Text Encoding Initiative). Il est normalisé ISO 12200:1999 Applications informatiques en terminologie -- Format de transfert de données terminologiques exploitables par la machine (MARTIF) -- Transfert négocié

[18] La norme ISO 12620:1999 fournit les catégories de données terminologiques qui servent à représenter les unités d'information d'un langage de représentation de données terminologiques. Chaque catégorie de données est modélisée par un ensemble de propriétés décrites à l'aide du modèle RDF (Resource Description Framework).

*important dès lors qu'il s'agit de permettre un échange des relations dites conceptuelles engrangées dans les bases de connaissances terminologiques* » (Romary & Campenhoudt, 2001). En effet, si l'on est tous convaincus du principe primordial selon lequel l'échange et l'intégration des données terminographiques constituent l'un des aspects majeurs du domaine de la terminologie, cet échange et cette intégration ne devraient pas se limiter aux simples listes des termes qui constituent un champ lexical ou un domaine conceptuel quelconque. La description et l'échange des relations de sens sont désormais au cœur même de l'aménagement apporté aux normes traditionnelles d'échange en terminologie. Défini sur la base de la norme ISO 12620:1999 qui fixe les catégories de données terminologiques dans une démarche d'échange, le modèle MARTIF part d'une sélection préalable (par négociation) de champs (éléments de données) communément utilisés dans les bases de données terminologiques. Ceci réduit inéluctablement l'importance du facteur d'interopérabilité entre elles.

TMF est concrètement développé dans l'optique d'élargir le canevas de ces échanges négociés vers des échanges « aveugles » entre des formats de références comme Martif, Geneter ([19]) et DXLT ([20]) ou d'autres à venir. L'idée était de définir une norme unique de description de formats de représentation et d'échange de terminologies informatisées. Cette description serait indépendante d'un choix explicite d'une implémentation particulière de cette norme. Elle pourrait partir d'une extension du méta modèle ISO 16642 à l'époque de la réalisation du CD (2000). Le principe retenu dans la forme actualisée de la norme ISO 16642:2003, est de définir une structure minimale commune à laquelle doivent se conformer tous les langages de représentation de données terminologiques (*TML : Terminological Markup Language*). Le rôle de TMF serait d'assurer la communication automatique entre deux langages de représentation de données par le moyen des filtres de transfert des formats vers une représentation abstraite intermédiaire nommée GMT (*Generic Mapping Tool*) (Fig.70). Ainsi, TMF ne représente pas *in fine* un format particulier mais plutôt un méta-modèle qui autorise la spécification des contraintes propres à un langage de description de données terminologiques, en l'occurrence le langage TML (*Terminology Mark-up Language*). Les différents langages TML compatibles, au TMF, sont interopérables entre eux et utilisent les mêmes catégories de données ([21]). L'avantage de cette approche est que l'ensemble des formats (ou TML) compatibles avec la plate-forme TMF constituerait une famille dont il est possible de définir rigoureusement les conditions d'interopérabilité. « *GMT, par son caractère abstrait, apparaît en définitive comme un intermédiaire idéal entre deux TML particuliers, notamment quand il s'agit de définir des filtres de l'un vers l'autre* ». (Romary, 2001).

---

[19] Geneter constitue l'Annexe C de la norme ISO 16642:2001 - Applications informatiques en terminologie -- Plate-forme pour le balisage de terminologies informatisées. C'est un format qui décrit le modèle d'une entrée terminologique dans une base de données terminologique. Le format Geneter prévoit trois types de catégories de données : un type de niveau supérieur (Top Level), un type de niveau intégré (Embedded Level) et un type de niveau de base (Basic Level). Chacune des catégories des données est décrite par trois types d'éléments : un Nom, Attributs et un Modèle de Contenu.

[20] Le but de DXLT est d'encoder en XML un ensemble de termes avec leurs définitions

[21] Une catégorie de données et l'ensemble d'éléments de données (ou champs) apparentés. Par exemple : la catégorie de données « Droit » du schéma LOM spécifie les conditions d'utilisation d'une ressource (coûts, droits d'auteur, etc.)

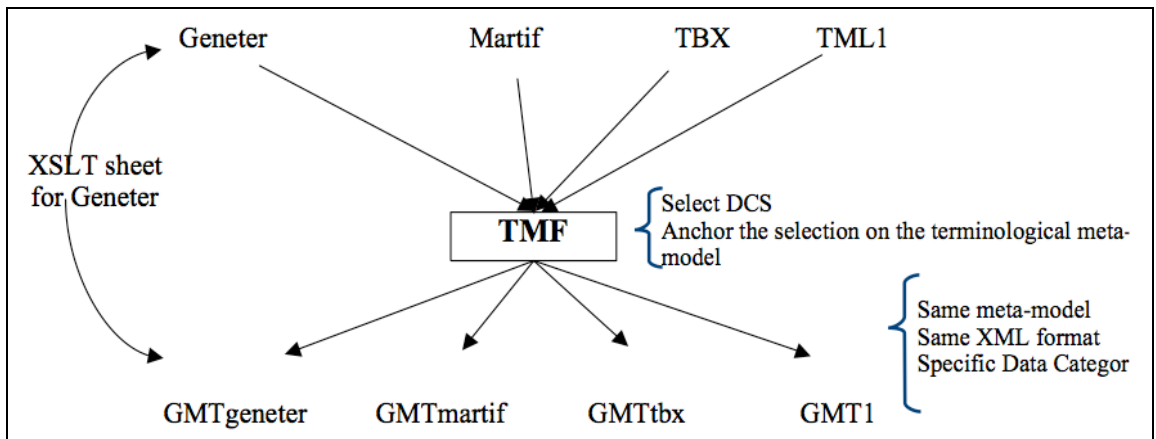

Fig. 4 : L'interopérabilité et les niveaux d'échanges par TMF (Source : Kramer, 2004)

Le modèle TMF répond à une structuration de données qui distingue trois niveaux hiérarchiques de description (Fig.4) : un niveau de données conceptuelles communes à toutes les langues, un niveau de données exclusives à une langue (réalisations linguistiques), et un niveau de données propres à un terme (informations lexicales). Il faudrait rappeler ici encore (nous l'avons expliqué comme argument en faveur de l'approche onomasiologique) qu'étant donné le nombre très élevé de termes dans un nombre très élevé de langues, la monosémie est, dans ce cas, la meilleure solution pour un meilleur contrôle de l'accroissement des contenus des bases de données terminologiques. Une telle organisation a permis la description des bases terminologiques par la combinaison de deux composantes :

1. Un métamodèle structurel (ou squelette) commun à tous les langages de représentation de données terminologiques composé de sept nœuds structurels, dont certains peuvent être répétitifs. Exemple : avoir plusieurs entrées linguistiques sous un nœud d'une entrée terminologique. La formule serait dans ce cas : un seul concept est décrit dans 'n' langues et désigné par 'n' termes dans chaque langue ('n' étant une valeur égale ou supérieure à 1);

2. Des informations complémentaires de description des rattachements aux différents nœuds du modèle structurel. Chaque description fait référence à une catégorie de données spécifiée dans la norme ISO 12620 ou définie de façon spécifique par le concepteur de la base terminologique.

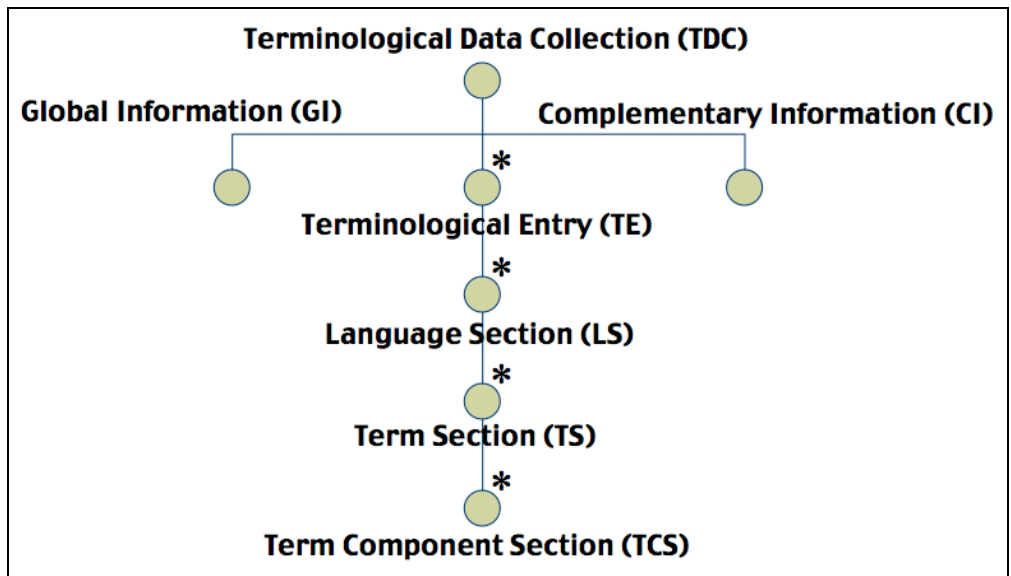

Fig. 5 : Les niveaux du métamodèle TMF (Source : Norme ISO 16642:2003)

Comment sont définis les sept nœuds de ce squelette de modèle TMF et comment s'articulent-ils entre eux ?

- Le nœud TDC (*Terminological data Collection*) : il s'agit de la collection de données contenant des informations sur les concepts dans un domaine spécifique. Ce niveau est celui de la base terminologique elle-même ;

- Le nœud GI (*Global Information*) et CI (Complementary information). Ce sont deux niveaux (en fait deux registres) dans lesquels sont stockées les données référentielles importantes pour administrer ou faire fonctionner la base. Ces données n'appartiennent pas directement à la collection terminologique.

- Le nœud TE (*Terminological Entry*) : C'est l'entrée qui contient les informations sur les unités terminologiques. Malgré son nom c'est exclusivement le niveau du concept. Certains experts s'accordent même pour convenir qu'à ce niveau l'adjectif « *conceptual* » aurait été plus clair que « *terminological* ». Tous les termes en toutes langues sont traités à un niveau hiérarchique inférieur lié à ce niveau de « l'entrée » et réfèrent tous à ce même identifiant. C'est aussi à ce niveau que l'on peut décrire les caractéristiques du système (ou graphe) de concept : générique ou partitif;

- Le nœud LS (*Language Section*) : C'est une section de l'entrée terminologique contenant des informations relatives à la langue. Comme son nom l'indique, c'est plus une section qu'un niveau. Le LS intégrera tout ce qui est dépendant des langages, et il s'oppose clairement au niveau des concepts. C'est à ce niveau hiérarchique que l'on ouvrira des langues. L'ouverture d'un LS est un préalable indispensable à l'ouverture d'un TL (*Term Level*) qui est un niveau encore hiérarchiquement inférieur dans chaque langue du schéma TMF ;

- Le nœud TL (*Term Level*) : C'est à ce niveau que s'ouvrent 1 ou 'n' termes supposés être tous plus ou moins synonymes dans autant de LS qu'il y a de langues dans la base. C'est à ce niveau que peut se faire une description morphosyntaxique des termes : genre, nombre, catégorie du discours (nom, adjectif, verbe, syntagme, mots valise..), abréviation, acronyme.

- Le nœud TCL (*Term Component Level*) : C'est à ce niveau que le modèle TMF permet de décrire (comme son nom l'indique) les composants d'un terme. Dans le cas concret du SC36 et dans beaucoup de terminologies d'instances normatives nous devons gérer une proportion importante de mots valises ou de syntagmes. Beaucoup de terminoticiens le considèrent comme un niveau de détail inutile. Formellement chaque élément d'un terme composé, dans chaque langue permet d'ouvrir un TCL (*Term Component Level*). Il est possible donc d'ouvrir autant d'item pour un terme qu'il y a de composants dans le terme. L'utilité réside en ce que ce TCL facilite une vision comparée (largement multilingue et potentiellement sémantiquement assistée) des modes de génération des nouveaux concepts d'un champ terminologique donné (néologisme ou néonymes). Cela permettra d'assister le travail des commissions nationales de terminologie techniques qui proposent et/ou cherchent à imposer des mots composées souvent très lointain les uns des autres dans des langues pourtant cousines comme le français et l'anglais.

À la suite de cette explication du métamodèle TMF, le schéma logique structurel qui permet d'optimiser la démarche onomasiologique s'organise dans les niveaux TE (*Terminological Entry*), LS (*Language Section*), TL (*Term Level*) et TCL (Term Component Level).

Dans notre approche de recherche-action, annoncée précédemment, cette particularité a été reprise dans notre initiative de production d'une terminologie multilingue et interopérable dans le domaine de l'éducation. C'est dans le cadre de nos engagements avec le groupe de travail WG1 du SC36, et dans les perspectives du projet de l'Alliance Cartago[22], point de liaison avec le WG1 pour la création de terminologie *e-Learning* multilingue, que nous avons pu produire un corpus terminologique en plusieurs langues. Cette production fondée essentiellement sur le modèle TMF, dépasse en volume et en nature de données de description, l'état actuel de la liste normalisée ISO 2382-36 et ses quelques dizaines de termes. Quelle démarche avons-nous suivi et quels objectifs estimons-nous avoir atteint à travers ce projet ?

## 3. Construire des terminologies multilingues normalisées dans le contexte de l'e-Learning et des TICE

Dans le monde de l'éducation, le développement de référentiels terminologiques concernant les technologies, les descriptions d'institutions, les disciplines, les modes de certification (diplômes, niveaux), les styles pédagogiques, les contextes juridiques etc. sont de plus en plus nécessaires pour permettre la circulation internationale des ressources d'enseignement et de formation. La généralisation constante des nouveaux modes d'enseignement par les TICE fait de l'enseignement, de la formation et de l'apprentissage des valeurs universelles de développement. La mise en place de ressources terminologiques multilingues, libres d'accès, qui soient un véritable « bien public », est, dès lors, indispensable dans un cadre de respect des langues utilisées dans le monde. Or, actuellement, une interopérabilité normalisée mondiale n'est assurée, partiellement, que par une seule langue : l'anglais. L'offre actuelle de terminologies est totalement inadéquate car seules quelques langues sont accessibles en raison de méthodes orientées sur les termes et non les concepts. Pourtant, de telles ressources sont utiles à tous ceux qui ont besoin de connaître, de façon fiable et relativement complète, l'état de l'offre en technologies et en programmes de formation électronique dans le monde. C'est bien là, une des attributions du SC36 ; celle de fournir non pas la totalité d'un corpus terminologique *e-Learning* multilingue, mais de mettre en application les procédures normalisées de conception et développement de ces corpus et leur appropriation par les experts internationaux, chacun son contexte, sa langue et sa culture. Comment définir alors le cadre général dans lequel agit le SC36 ? Dans quelle perspective répond-il à une demande concrète des pays membres et du monde entier ([23]) ? Quels sont ses outils, ses choix et ses prérogatives pour y parvenir ?

---

[22] Fondé à Tunis (d'où son nom) lors du SMSI Tunis 2005 dans le cadre d'Initiative 2005, colloque commun de l'AUF et du SC36.

## 3.1. Les chantiers terminologiques en normalisation e-Learning et TICE : stratégies et raisons d'être

Le SC36 s'est ainsi conformé aux règles de la majorité des instances normatives et il a crée un groupe de travail : le SC36-WG1 : terminologie qui se restreint assez vite pour devenir un groupe vocabulaire. Il s'avère que pour des raisons diverses la normalisation de l'e-learning a surtout été amorcée par les « pionniers du domaines » essentiellement des industriels de l'e-training aéronautique, militaire et de sécurité auxquels se sont joint des enseignants européens des sciences exactes et expérimentales. Ces pionniers s'accommodent incomparablement mieux que les nouveaux entrants actuels d'un unilinguisme pragmatique (c'est le cas quasi à 100% dans le secteur aéronautique), d'une relative univocité des concepts des matières enseignées et d'autre part ce sont des domaines pédagogiques dans lesquels les débats sur la diversité des cultures pédagogiques sont moins présents que dans les arts ou les lettres.

Cette première prépondérance des pionniers s'estompe aujourd'hui (même s'ils représentent toujours une part importante du marché global), les cultures linguistiques autres qu'anglophones et européennes (notamment Corée, Chine et Japon) affirment d'année en année un dynamisme et une croissance importante. Le Sud-est asiatique pourrait à lui seul multiplier le marché de façon considérable, et pourtant ce n'est que très récemment que cette partie du monde commence à affirmer son droit à la diversité linguistique et culturelle dans le processus normatif ([24]).

L'évolution historique du SC36 qu'il serait hors sujet de relater en détail fait que nous disposons maintenant d'un groupe de travail produisant un vocabulaire à ce jour bilingue (anglais/français), co-élaborée en consensus dans les deux langues. Ce

---

[23] La distinction est importante parce que par exemple les 2 auteurs sont à la fois délégués du NB France (AFNOR) en tant que représentants de deux universités françaises (Paris 8 et Bordeaux 3, mais aussi l'université La Manouba en Tunisie). Ils sont aussi « Liaison AUF », donc représentants des intérêts de la diversité des langues de la Francophonie. Ils sont aussi « Liaison Cartago » donc responsables du respect d'une orthodoxie de méthode en terminologies normatives chargées aussi du respect de diversité mondiale des langues et cultures.

[24] Voir Henri Hudrisier, « Société de connaissance, le paradigme de l'appropriation », ou encore Mokhtar Ben Henda, « Les contradictions d'une politique de diversité culturelle », in Hermès 45 « *Fractures dans la société de connaissances* », Sous la dir. de Didier Oillo et Bonaventure Mvé-Ondo. CNRS éditions, Paris 2006

vocabulaire n'a pas vocation à être très important (quelques centaines). Par décision de l'instance hiérarchiquement supérieure au SC36 (le JTC1) ce vocabulaire devrait en principe être mis en commun avec tous les autres vocabulaires ou terminologies du JTC1 dans une base terminologique commune gérée par Termium ([25]).

De nombreux problèmes restent ouverts :

Les experts des TICE ne sont pas forcément sensibilisés aux problèmes méthodologiques de la terminologie et de la terminotique. Il a été long et difficile pour les quelques délégués du SC36 qui étaient sensibilisés aux méthodes terminologiques de faire comprendre au groupe de travail vocabulaire (SC36-WG1) que les termes anglais n'étaient pas équivalent à des concepts universaux

Il a été aussi difficile (face au groupe des professionnels et animateurs de la normalisation[26]) de faire comprendre que contrairement à la plupart des autres normes qui portaient jusqu'à ces dernières années sur des produits ou des services très techniques et très matériels, le SC36 appartenait (et même était emblématique) des nouveaux champs de normalisation beaucoup plus culturels et sociétaux.

D'évidence les normes des TICE achoppent sur des domaines très politiques, sur des options éthiques, sur le champ du linguistique[27]… C'est pourquoi les experts en normalisation (notamment les professionnels des institutions de normalisation) ont du mal à accompagner (voire à comprendre) cette mutation évolutive de la normalisation des produits matériels vers la normalisation de services sociétaux. Cette évolution est au cœur des questions posées par la gouvernance mondiale.

La production de ces normes de service sociétaux exige (non plus la mise en place de terminologies sous forme de nomenclatures techniques[28] mais de véritable terminologies (voire d'ontologies) beaucoup plus vastes et sophistiquées, référent à la diversité des cultures, des langues, des nations sans pour autant perdre de vue que tout projet normatif mondial exige des référentiels sémantiques universels permettant précisément l'interopérabilité des réseaux ou des échanges de ressources.

---

[25] C'est la base de données terminologique Termium du Bureau fédéral de la traduction canadien. Celle-ci a offert ses services au JTC1 pour collecter et devenir le lieu de référence de la terminologie de tous les sous-comités qui en dépendent (y compris la terminologie e-learning du SC36). La base Termium, qui gère un fonds très important de ressources terminologiques est cependant organisée prioritairement pour répondre aux besoins de traduction d'un pays bilingue. De ce fait elle propose aujourd'hui des terminologies en français et anglais. A moyen terme elle est prévue pour être opérationnelle en 4, ou 5 langues au plus

[26] L'ensemble des personnes qui constituent un groupe de normalisation (par exemple le SC36) se distribuent en différentes catégories. Des experts du domaine (non obligatoirement formés à la normalisation, par ex. au S36, des pédagogues, des industriels informatiques, des éditeurs d'e-enseignement), des experts liaison (un expert MPEG, un expert de l'e-learning aéronautique…), mais aussi des professionnels employés par les instances de normalisation (ISO, AFNOR, BSI…). Ce sont ces professionnels de la normalisation ainsi que les experts les plus dynamiques et les plus expérimentés (qui ont participés auparavant à d'autres domaines de normalisation) qui constituent bien sûr le noyau dirigeant permettant la production effective de la famille de normes que doit éditer l'instance (ici le SC36)

[27] On n'apprend pas selon la même logique une langue alphabétique ou une langue idéographique ; et on sait que l'enseignement de la langue et de l'écriture sont aux fondements même de la transmission de savoir

[28] Les normes plus strictement techniques (métallurgie, composants électroniques, etc.) posent très peu de problèmes de découpage conceptuel dans la diversité des langues. Les terminologies de ces domaines techniques s'accommodent dès lors très bien de la réalisation de listes de vocabulaire en anglais qu'il suffit de traduire terme à terme dans d'autres langues

On comprend bien en effet que la terminologie du SC36 appartient à l'évidence, pour une part, à des nomenclatures techniques relevant du premier cas (système de gestion d'apprentissage, environnement d'apprentissage, métadonnées pour des objets d'apprentissage, système de gestion de contenu, environnement virtuel d'apprentissage…). D'évidence, ces notions ne posent pratiquement aucun problème quand à leur découpage dans la diversité des langues : les TIC se sont construites presque partout selon le même découpage conceptuel[29].

Par contre, le problème se complique quand on s'attaque aux termes généraux, *apprentissage*, *éducation*, *formation* ou même des termes comme *e-learning*, *blended learning* qui posent des problèmes de diversité de traduction et surtout d'appréhension conceptuelle très délicat. L'interopérabilité entre terme devient inextricable quand on décrit un apprenant ou un enseignant et ses rôles, des institutions éducatives, des niveaux d'apprentissage, des modes de diplomation.

D'évidence il ne peut être question d'envisager que l'ISO, l'IEC, le JTC1 se substituent à l'UNESCO ou aux grandes agences mondiales impliquées dans la culture ou l'éducation pour ce qui est de la réalisation de grandes banques terminologiques ou de vastes terminologies qui serait susceptible d'accompagner l'organisation normative (notamment la capacité d'interopérabilité donc de circulation équitable) des ressources et des acteurs des TICE. Ces enjeux sont stratégiques on le sait, mais ce n'est pas le rôle direct des instances de normalisation. Il ne peut être de la responsabilité directe du SC36-WG1 de réaliser des terminologies de plusieurs milliers de termes dans un nombre extensif de langues pour assurer l'interopérabilité normative de ce domaine technologique.

Après que le problème ait été longuement discuté, un nombre significatif d'experts sont maintenant persuadés que seule une approche privilégiant les concepts définis en consensus vers lesquels « pointent » des termes organisés eux-mêmes chacun dans leur langue, peut clarifier la question. C'est la méthode onomasiologique, informatisable grâce au savoir faire terminotique ; c'est tout cela qui a été normalisé à l'ISO TC37 grâce à l'effort de plusieurs décennies de travail par les normalisateurs terminologues et terminoticiens.

Cependant le SC36-WG1 continue sa première démarche en faisant progresser une liste bilingue de vocabulaire. La première édition (en cours de normalisation définitive sous la référence ISO/IEC 2382-36) fait déjà place au chantier d'une seconde édition augmentée. Cette démarche est de notre point de vue parfaitement légitime à condition (ce qui est le cas) que le SC36 envisage de collaborer en synergie avec une instance qui prenne le problème selon des méthodologies plus adaptées à l'aspect sociétal et culturel que constituent les TICE. C'est ce qui est actuellement en cours de mise en place notamment avec l'aide de l'Alliance Cartago.

---

[29] A l'exception notable de certaines langues (minoritaires ou dominées) qui n'ont pas eu l'opportunité socio-éducative d'accompagner la modernité et qui ont été victimes de perte de domaines linguistiques plus ou moins vastes. Nous y reviendrons mais un des objectifs éthiques de Cartago est de tenter de freiner, voire de réparer pour partie ces déficits, dans un champ sémantique aussi stratégique que la transmission de savoir.

# Conclusion

Réaliser des terminologies normalisées multilingues est l'une des premières actions amorcées dans nos activités normatives au sein du SC36. Les instances de normalisation constituent de façon quasi systématique une terminologie de leur domaine. Dans les domaines très techniques (composants informatique par exemple) un « vocabulaire unilingue anglophone » répond assez souvent aux besoins des utilisateurs constructeurs dans un domaine de normalisation comme les TICE. On comprend dès lors que les besoins sont beaucoup plus liés aux spécificités linguistiques ; et c'est là tout l'enjeu des communautés non anglophones qui doivent pouvoir exprimer leur spécificité linguistique dans la définition de la matière académique (disciplines), des spécificités de formation (métiers), des spécificités institutionnelles (organisation des niveaux académiques, des acteurs et des institutions scolaires et universitaires, de l'organisation nationale de la formation, etc.). C'est la raison pour laquelle nous sommes engagés dans ce chantier normatif afin de créer la jonction de notre domaine de recherche sur les systèmes d'information et de communication avec le champ pratique innovateur de la gestion et la création de ressources et de contenus terminologiques mondiaux et largement multilingues dans le domaine spécialisé de l'e-Learning et des TICE.

Après vous avoir présenté à un niveau théorique les tenants et les aboutissants de notre approche de travail, nous tenons à inviter un maximum d'intéressés s'associer à notre démarche de construction de terminologies multilingues normalisées. Si nous sommes ici dans ce colloque organisé à Sousse, c'est par envi de nous adresser à un auditoire académique arabophone et multilingue, déjà acquis à l'importance vitale de ce domaine terminologique pour la science et la recherche. Les trois dimensions principales du multilinguisme du Monde arabe (arabe, anglais, français) sont extrêmement pertinentes pour notre propos et elles sont indispensables pour que ce même Monde arabe (et ses différentes traditions dialectales) puisse pouvoir continuer à étudier et à produire de la science et de l'industrie en arabe. Le meilleur remède contre la perte de domaines lexicaux est en effet de contribuer continument à des travaux terminologiques multilingues reliés à sa propre tradition linguistique maternelle et aux langues dominantes du développement industriel. En l'occurrence l'ISO SC36 a choisi de co-élaborer de façon normalisée une terminologie des TICE référentiellement en anglais et en français. Il est indispensable qu'une collaboration pan arabique du monde savant de l'éducation contribue à y élaborer et à y maintenir une terminologie arabe exhaustive et cohérente. Nous avons commencé à élaborer les premières ébauches d'une terminologies en plusieurs langues (dont l'arabe). Il nous reste la consolidation de nos réalisations par la contribution d'experts terminologues dans les différentes langues partenaires. Pour la langue arabe, nous avons en ligne de mire les institutions officielles comme le bureau de coordination de la langue arabe de l'Alecso, mais nous comptons aussi sur la contribution de la dynamique académique pour des affinités de recherche-action et de productivité de valeur scientifique experte. Nous fournissons (en annexe de cette communication) les principales normes ou projets de normes qui servent de cahier des charges à ce travail.

La morale dans tout ce que nous avons exposé est qu'il est banal de souligner que le langage est au cœur de la capacité humaine d'apprendre et de s'adapter. Il s'agit dès lors de permettre à l'espèce humaine de préserver sa diversité linguistique, donc son potentiel adaptatif tout en assumant l'évidence indispensable de la globalisation multilingue. Mais comme bien d'autres évidences écologiques, ces chantiers d'aménagement numérique équitable de toutes les langues dans leur diversité lexicales, syntaxique et scripturale seront-ils pris en compte ? Continuerons-nous à pouvoir apprendre, travailler, commercer, nous distraire dans la diversité de toutes nos langues maternelles et contradictoirement à vivre harmonieusement la globalisation de plusieurs milliards d'hommes sur une si petite planète ? Ces chantiers de développement d'une traductique, d'une terminotique, de TIC et de TICE véritablement interopérables, et communiquants parce que normalisés sont d'année en année plus urgents et plus indispensables pour l'avenir de toutes les communautés culturelles et linguistiques en particulier, mais aussi pour la survie même de l'espèce humaine.

# Bibliographie